\documentclass[aps,preprint,nofootinbib,superscriptaddress]{revtex4}
\usepackage{graphicx,color}
\usepackage{amsmath,amssymb,amscd,amsfonts}
\usepackage{subfigure}
\usepackage{slashed}
\usepackage{bm}
\usepackage{xcolor}
\usepackage{diagbox,pict2e}

\def \Lag{\mathcal L}

\def \ord{\mathcal O}
\newcommand{\bwt}{\begin{widetext}}
\newcommand{\ewt}{\end{widetext}}
\newcommand{\newc}{\newcommand}
\newc{\hc}{\dagger}
\newc{\pd}{\partial}
\newc{\beq}{\begin{equation}}
\newc{\eeq}{\end{equation}}
\newc{\beqa}{\begin{eqnarray}}
\newc{\eeqa}{\end{eqnarray}}
\newc{\bi}{\begin{itemize}}
\newc{\ei}{\end{itemize}}
\newc{\ra}{\rightarrow}
\newc{\la}{\leftarrow}
\newc{\lra}{\longrightarrow}
\newc{\lla}{\longleftarrow}
\newc{\Lra}{\Longrightarrow}
\newc{\Lla}{\Longleftarrow}
\newc{\half}{\frac{1}{2}}
\newc{\fth}{\frac{1}{4}}
\newc{\del}{\delta}
\newc{\Del}{\Delta}
\newc{\gm}{\gamma}
\newc{\Gm}{\Gamma}
\newc{\lam}{\lambda}
\newc{\kap}{\kappa}
\newc{\tri}{\triangle}
\newc{\eps}{\epsilon}
\newc{\epsp}{\epsilon^\prime}
\newc{\wt}{\widetilde}
\newc{\ovl}{\overline}
\newc{\tchi}{\tilde{\chi}}
\newc{\ds}{\displaystyle}
\newc{\pmt}{\pm\!\pm}
\newc{\PL}{\hat{L}}
\newc{\PR}{\hat{R}}
\newc{\st}{s_\theta}
\newc{\ct}{c_\theta}

\newc{\msm}{\mathrm{SM}}
\newc{\msh}{\mathrm{sh}}
\newc{\mtev}{\mathrm{TeV}}
\newc{\mgev}{\mathrm{GeV}}
\newc{\mmev}{\mathrm{MeV}}
\newc{\mkev}{\mathrm{keV}}
\newc{\mev}{\mathrm{eV}}
\newc{\Tr}{\mathrm{Tr}}
\newc{\non}{\nonumber}
\newc{\clbl}{\color{blue}}
\newc{\clg}{\color{green}}
\newc{\clr}{\color{red}}
\mathchardef\mhyphen="2D
\newc{\SL}{\not\!\!}

\begin{document}
\title{KeV scale new fermion from a hidden sector}
\date{\today}

\author{We-Fu Chang}
\email{wfchang@phys.nthu.edu.tw}
\affiliation{Department of Physics, National Tsing Hua University, No. 101, Section 2, Kuang-Fu Road, Hsinchu, Taiwan 30013, R.O.C.}
\author{John N. Ng}
\email{misery@triumf.ca}
\affiliation{TRIUMF, 4004 Wesbrook Mall, Vancouver, BC, V6T 2A3, Canada}
\begin{abstract}
We studied a simple model of hidden sector that consists of a Dirac fermion $\chi$ and a spontaneously broken $U(1)_s$
symmetry. The dark sector is connected to the Standard Model(SM) via three righthanded SM singlet neutrinos, $N_R$'s, and the
kinetic mixing between $U(1)_s$ and $U(1)_Y$. A mixing between the scalar $\phi$ that breaks $U(1)_s$ and the SM Higgs boson, $H$, is implemented via the term $\phi^\dagger \phi H^\dagger H$ and this provides a third connection to the SM. Integrating out the $N_R$ at a high scale not only gives the active neutrinos, $\nu$, masses but generates effective Dirac-type couplings between $\nu$ and $\chi$.
This changes the usual Type-I seesaw results for active neutrino masses and makes $\chi$ behave like a sterile neutrino even though its origin is in the hidden sector. Note that $\chi$ is also split into a pair of Majorana
fermions. The amount of splitting depends on the parameters. If the lighter of the pair has a mass around keV, its lifetime is
longer than the age of the Universe and it can be a warm dark matter candidate. Signatures of
$\chi$ in high precision Kurie plots of nuclei $\beta$ decays and low energy neutrino nuclei coherent
scatterings are discussed. The model also induces new invisible $Z$ decay modes that can be searched
for in future Z factories.

\end{abstract}
\maketitle
\section{Introduction}

Some time ago we investigated a very simple shadow $U(1)_s$ sector which consists of a scalar $\phi$ that spontaneously breaks the gauged Abelian symmetry \cite{CNW1}.
The resulting massive gauge boson $X^\mu$ was allowed to kinetically mix with the hypercharge
gauge boson $B^\mu$ of the Standard Model (SM).
The scalar $\phi$ also couples to the SM Higgs field $H$  via the term $\phi^\dagger \phi H^\dagger H$. In today's parlance, this will
be the simplest two portal model respecting the SM gauge symmetry. The first portal is a vector one with the hypercharge as the mediator and the second is a scalar portal mediated by the Higgs boson. $U(1)_s$ symmetry breaking scale characterized by $v_s$ is taken to be above the electroweak breaking scale given by $v= 247 \mgev$.  A scale-invariant version for the scalar sector was also constructed \cite{CNW2}.
 Moreover, the models did not yield a dark matter candidate. In this paper, we extend the model by adding a massive Dirac fermion $\chi$ which is a SM singlet but is charged under the local $U(1)_s$. We also included at least two heavy righthanded SM singlet neutrinos, $N_R$, which are singlets under $U(1)_s$. Doing so enables us to use the type-I seesaw mechanism for active neutrino masses.
 For clarity's sake, much of our discussion will be given for one $N_R$ and extending to the realistic case of 3 $N_R$ is straightforward.
Here, $N_R$ will also play the dual role of a fermion portal to the hidden sector. This constitutes a  very simple complete minimal model with all three portals present.

 Since the physics of the $U(1)_s$ gauge boson and the scalar has been discussed thoroughly before,  we  concentrate here on the fermion $\chi$.
 In particular, we explore the parameter space which allows $\chi$ to be a dark matter candidate. Using the minimal content of the hidden sector and the conventional breaking of the $U(1)_s$  does not leave us with a symmetry that can protect $\chi$ from decaying; thus, in general, it cannot be stable. Without imposing an ad hoc symmetry, the only open option is to arrange $\chi$ to be long-lived, and it plays the role of warm dark matter (WDM); similar to that of a sterile neutrino. Recently, WDM receives increasing attention due to its ability to address the small scale problem of the cold dark matter plus cosmological constant ($\Lambda$CDM) paradigm. Note that $\Lambda$CDMs have DM masses in the GeV to TeV range,  and they predict too many satellite galaxies in the Milky Way and cusped DM profiles which contradicts current observations.
 On the other hand, DM with masses in the keV range are capable of accommodating the number of observed
 satellites as well as cored profiles of dwarf galaxies which are believed to be DM dominated.

  The satellite problem arises because free relativistic particles do not cluster and they erase
  structures of scale smaller than the particle free-streaming length $\ell_{\mathrm{fs}}$ which is
  approximately the distance traveled before the particle becomes non-relativistic $\sim c/3$.
  For the keV scale $\ell_{\mathrm{fs}}\sim 100 {\mathrm{kpc}}$. On the contrary, for cold dark matter ( CDM ) which is heavier
  and slower, has $\ell_{\mathrm{fs}}$ that are a million times smaller, resulting in too many small scale structures
  \cite{KT}. While CDM is very successful in accounting for large scale structures and many other cosmological observations (see \cite{BBK} for a review), it gives a steep cusp at the center for the galaxy density profile $\rho \sim r^{-1}$ \cite{NFW}. In contrast, WDM gives a finite constant density at the center $\rho \sim \rho_0$ which is more in line with observations \cite{deB}.

  In addition, there are claims of the detection of a monochromatic line at 3.56 $\mkev$ X-ray data towards the Andromeda and Perseus cluster \cite{Bul}and \cite{BRIF}. This can be interpreted as the radiative decay of a fermion, usually taken to be a sterile neutrino, into an active neutrino plus a photon. While this is suggestive, a more mundane astrophysical explanation is also possible.
 Here we explore the possibility that monochromatic gamma can come from the radiative decay of $\chi\ra \nu +\gm$ for a range of masses of interest in explaining the small scale structure conundrum.
   This motivates us to focus the $\chi$ mass in the range of $2- 10 \mkev$. We shall see later that the astrophysical and cosmological properties of $\chi$ we arrived at is almost indistinguishable from those of a sterile neutrino. A lucid review of sterile neutrinos as warm dark matter can be found in \cite{snudm}.
    As expected, if $\chi$ were a viable WDM candidate, the parameter space of the model will be restricted.
We emphasize that $\chi$ is conceptually and physically different from a sterile neutrino since it is not connected to active neutrino mass generation.
 Moreover, the $\chi$ fermions in our model come in as a vector pair. If they acquire a sizable mass splitting, the lighter one still serves as the WDM while the heavier one is much less restricted than the keV sterile neutrino WDM, and could have low energy phenomena. Exploration of this is one of the purposes of this paper.

   The paper is organized as follows. In Sec.\ref{sec:2},  we discuss in detail the model and the role of the high scale type-I seesaw. This leads to the lifetime of $\chi$ in Sec.\ref{sec:3}.
 The implications of $\chi$ for low energy precision neutrino physics are given in Sec.\ref{sec:4}. Effects on $\beta$ decays of nuclei, neutrinoless double beta decays of nuclei and recently observed coherent low energy neutrino scattering producing $\chi$ will be examined. Next, we give miscellaneous considerations of $\chi$ at higher energies in Sec.\ref{sec:5}. The main new result is the additional invisible decay of the SM $Z$.  In Sec.\ref{sec:6}, we discuss the cosmological requirements of  $\chi$ as the viable WDM and the limits they set on the parameters of the model.  Sec.\ref{sec:7} contains our conclusions.

 \section{ Type I seesaw and the three portals to the hidden world}
 \label{sec:2}
    In the usual notation, the gauge group of the
  model is $SU(2)_L\times U(1)_Y \times U(1)_s$
  where the color sector is omitted.  We begin by discussing only one generation. The fields  beyond the SM ones we use are the SM singlet righthanded neutrino, $N_R$, the hidden Dirac fermion $\chi$ which can be considered as a pair of different chirality Weyl fermions $\chi_{L,R}$, a hidden sector scalar, $\phi$, and the gauge field $X_\mu$  of $U(1)_s$. The fields and their relevant quantum numbers are given in Table I.

  \begin{table}
\begin{center}
\renewcommand{\arraystretch}{1.30}
\begin{tabular}{|c|c|c|c|c|c|c|}
\hline
Field&\;\;$\ell_L$\;\;&\;\;$H\;\;$&$\;\,N_R\;\,$ &$\;\;\chi_L\;\;$&\;\; $\chi_R\;\;$&\;\; $\phi$\;\;\\
\hline
$SU(2)_L$& 2&2 &1&1&1&1\\
\hline
$U(1)_Y$&-$\half$&$\half$&0&0&0&0\\
\hline
$U(1)_s$&0&0&0&1&1&1\\
\hline
\end{tabular}
\caption{$U(1)_s$ and SM quantum numbers for relevant fields.}
\label{tb:I}
\end{center}
\end{table}

The complete gauge invariant Lagrangian is given by
\beqa
\label{eq:LT}
\Lag&=& \Lag_{\msm I}+\Lag_{\msh}+\Lag_{N\chi}\,, \non \\
\Lag_{\msm I}&=& \Lag_{\msm} +  \ovl{N_R} i\slashed{\pd} N_R -\left( y\bar{\ell_L} N_R \tilde{H} +\frac{1}{2}M_N \ovl{N^{c}_R} N_R + h.c.\right)\,, \non \\
\Lag_{\msh}&=& -\frac{1}{4} X^{\mu\nu} X_{\mu\nu} -\frac{\eps}{2} B^{\mu\nu}X_{\mu\nu} +\biggl|\biggl(\pd_\mu - i g_s X_\mu\biggr)\phi \biggr|^2 +\bar{\chi} (i\slashed{\pd}-g_s \slashed{X}) \chi -M_\chi \bar{\chi}\chi -V(\phi,H)\,, \non \\
\Lag_{N\chi}&=& -f_L \ovl{\chi_L}\,N_R \phi -f_R \ovl{\chi^{c}_R}\,N_R \phi^* +h.c.\,, \non \\
V(H,\phi)&=&  - \mu_s^2 \phi^*\phi +\lam_s(\phi^*\phi)^2 +\kappa (H^\dagger H)(\phi^*\phi)  - \mu^2 H^\dagger H +\lam(H^\dagger H)^2\,,
\eeqa
where $\ell_L$ and $H$ are the SM lepton doublet and the  Higgs field, respectively. Also, $\tilde{H}=i\sigma_2 H^*$ as in the standard notation. Note that $B_{\mu\nu}$ is the field strength tensor of $U(1)_Y$, and $g_s$ is the gauge coupling of $U(1)_s$. We have arbitrarily chosen $\chi$ and $\phi$ to have unit shadow charge, with the convention given in Eq.(\ref{eq:LT}).
 It is interesting to note that a conventional lepton number of plus/minus one unit can be assigned to $\chi_{L/R}$ and 0 for $\phi$.
But this is unnecessary since the Majorana mass term for $N_R$ breaks it  explicitly by two units. The three portal terms are $\eps$ in $\Lag_{\msh}$, $\Lag_{N\chi}$, and the $\kappa$ term in $V(H,\phi)$.

\begin{figure}[h!]
\centering
\includegraphics[width=0.8\textwidth]{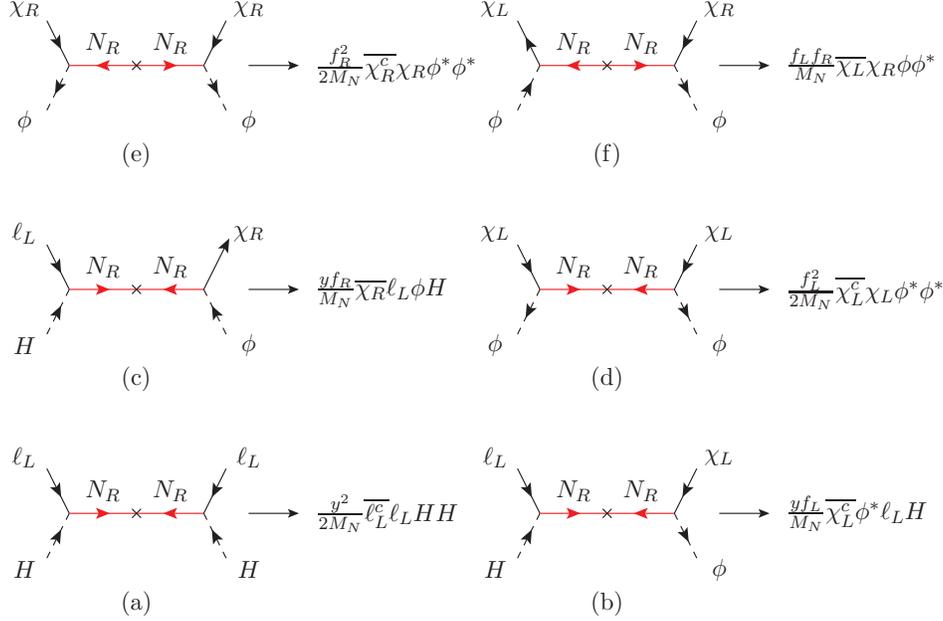}
\caption{ The effective Lagrangian $\Lag_5$ terms, after integrating out the heavy $N_R$, and the corresponding Feynman diagrams. The crosses represent the Majorana mass of $N_R$.}
\label{fig:seesaw}
\end{figure}

Next, we implement the type-I seesaw mechanism. Namely, we assume that $M_N$ is much heavier than any other masses, and we integrate out $N_R$. The easiest way to do this is diagrammatically.
As depicted in the Feynman diagrams given in Fig.\ref{fig:seesaw}, some dimension-five terms are generated below the scale $M_N$.

The effective theory below the seesaw scale then consists of $\Lag_{\msm}+\Lag_{sh}$ and a lepton-number-violating $\Lag_5$
\beq
\begin{split}
-M_N\Lag_5=&\frac12 y^2 \ovl{\ell_L^c} \ell_L H H +y f_L\ovl{\chi_L^c}  \phi^* \ell_L H +
y f_R \ovl{\chi_R} \ell_L \phi H \\
&+ \frac12 f_L^2 \ovl{\chi_L^c}\chi_L \phi^*\phi^* +  \frac12 f_R^2 \ovl{\chi_R^c}\chi_R \phi^*\phi^*
+ f_L f_R \ovl{\chi_L}\chi_R \phi \phi^* +h.c.
\end{split}
\label{eq:Lag5}
\eeq

After spontaneous symmetry breaking (SSB) with $\langle H \rangle=v/\sqrt{2}$, one gets the familiar Weinberg operator \cite{WO} for active neutrinos with mass $\sim \frac{y^2v^2}{2 M_N}$, see Fig. (1a). Besides, we also have SSB for $U(1)_s$ via $\langle \phi \rangle = v_s/\sqrt{2}$. Figs.(1b) and (1c) induce mass mixings with the hidden fermions given by
$\sim \frac{yf_{L/R}v v_s}{2 M_N}$. Repeating this for all the diagrams of Fig.(\ref{fig:seesaw}) yields, for one SM generation, a $3\times 3$  effective neutrino mass matrix $M_\nu$. In the weak basis $\{\nu, \chi_L,\chi_R^c\}$,  it is given by
 \beq
 \label{eq:numm}
M_\nu \sim \begin{pmatrix}y^2\frac{v^2}{M_N}&yf_L \frac{v v_s}{M_N} & y f_R \frac{v v_s}{ M_N} \\
 yf_L \frac{v v_s}{ M_N}& f_L^2 \frac{v_s^2}{M_N} &  M_\chi+ f_L f_R \frac{v_s^2}{M_N}\\
 yf_R \frac{v v_s}{ M_N}& M_\chi+f_Lf_R \frac{v_s^2}{M_N} & f_R^2 \frac{v_s^2}{M_N}
 \end{pmatrix}.
\eeq
We reiterate that $M_\nu$ is a consequence of generalizing the high scale type-I seesaw mechanism to include the fermion portal Lagrangian with SSB taken afterward. Thus, the hierarchy of the scale we are interested in is $M_N \ggg v_s \gtrsim  v$.
It is easy to see that with this hierarchy, the entry of active neutrino masses, i.e., the uppermost left corner of Eq.(\ref{eq:numm}), is in the sub-eV range for $\frac{y^2}{M_N}\simeq 10^{-14} (GeV)^{-1}$.
We use the benchmark point of $M_N=10^{10}\,\mgev$,  $y=10^{-2}$ and $v_s = 1\, \mtev$.
Thus, $\frac{v_s^2}{M_N}\simeq  100 \mkev$ and $\frac{v v_S}{M_N}\simeq 10 \mkev$. We also work in the perturbative regime and hence take $f_{L/R}\lesssim 1$.
Specifically  if we take $M_\chi \sim 10 \mkev$ and $f_L=f_R=0.1$,  we have
\beq
 \label{eq:numm1}
M_\nu \simeq\begin{pmatrix}y^2\frac{v^2}{M_N}&yf_L \frac{v v_s}{ M_N} & y f_R \frac{v v_s}{ M_N} \\
 yf_L \frac{v v_s}{M_N}& f_L^2 \frac{v_s^2}{M_N} & M_\chi\\
 yf_R \frac{v v_s}{M_N}& M_\chi & f_R^2 \frac{v_s^2}{M_N}
 \end{pmatrix}.
\eeq
It is easy to see that the splitting of $\chi_L$ and $\chi_R$ arises from the $f^2_{L, R}$ terms. For $f_{L/R}\sim 0.1$ this is $\ord (1 \mkev)$. Thus, $\chi$ will remain essentially a Dirac fermion for this range of parameters.  As a reminder, $M_\chi$ is not the physical mass. Even smaller splitting can be
obtained by taking  $f_{L/R}\ll 1$ . In that case, the physical states are actually two Majorana neutrinos $\chi_1,\chi_2$ with masses so close to each other that most experiments cannot resolve them.

For larger splitting we have to explore a different parameter region. If we take  $f_L=f_R=f\sim 1$, then $\frac{f^2v_s^2}{M_N}\sim 100 \mkev$. For $M_\chi \sim 80 \mkev$, the original Dirac $\chi$
will now split into two Majorana fermions one with mass $\sim 100 \mkev$ and the other with mass $\sim 10 \mkev$.
Another example is to take $M_N=10^8\, \mgev$ with  $y=10^{-3} $ so that the active neutrino masses will be in the sub-eV range via Type-I seesaw. With $f\simeq 0.3$, $M_\chi\sim \mmev$, this yields a mass at around $10\mkev$, with the other at about $\mmev$. From this exercise, it is clear to see that the mass splitting becomes more substantial if the lepton number violation scale $M_N$ is lower and/or $v_s, M_\chi $ are raised. Moreover, only the lighter one can be the WDM to solve the small scale problem of CDM. Nevertheless,
the more massive partner can have interesting phenomenology as we shall see later.

To make the physics more transparent for how the above considerations can alter the type-I seesaw mechanism for active neutrino mass generation and the value of the eventual physical mass of $ \chi$, we first consider the case where $f_L=f_R=f$, $y=1$, $M_N=10^{14}$GeV, and ignore the splitting discussed above.
The mixing of active $\nu$ is with a Dirac shadow fermion $\chi$. The simplified neutral fermion mass matrix $M_\nu^s$ becomes
\beq
\label{eq:mnus}
M_\nu^s \simeq\begin{pmatrix}\frac{v^2}{M_N}&\frac{fvv_s}{M_N}\\\frac{fvv_s}{ M_N} & M_\chi \end{pmatrix}.
\eeq
The eigenvalues are
\beq
M_0^{\pm} =\frac{M_\chi}{2}\left[(1+a)\pm\sqrt{(1-a)^2+4b^2}\right]\,,
\eeq
where $a= \frac{v^2}{M_\chi M_N}$ and $ b=\frac{fvv_s}{ M_\chi M_N}$ and $a,b\ll 1$ for
 $M_\chi> 100\mev$. Thus,
\beqa
\label{eq:M+}
M_2\equiv M_0^{+} &\simeq &M_\chi\left[ 1+   \left(\frac{f v v_s}{ M_\chi M_N}\right)^2\right]\,,  \\
M_1\equiv M_0^{-}&\simeq &\frac{v^2}{M_N}\left[ 1- \left(\frac{f^2v_s^{2}}{ M_\chi M_N}\right)\right]\,.
\label{eq:M-}
\eeqa
 The physical mass of $\chi$ is pushed up, but it does not change by much. On the other hand, the physical mass of the active neutrino is pushed down compared to the type-I seesaw value.

It is instructive to look at some typical numbers. Note that $M_\chi$ is a free parameter. If we take $v_s=1\mtev, M_N=10^{14}\mgev$ (here we set $y=1$), the correction to $M_\chi=1 \mkev$ is negligibly small.
In contrast, the correction to the seesaw active neutrino mass is $\sim f^2\,\%$ (see. Eq.(\ref{eq:M-})), which can be substantial if $f>1$.
Furthermore, the correction is more significant for smaller values of $M_\chi$.

The mixing angle is given by
\beq
\label{eq:chinumix}
 \theta \simeq  \frac{y f v v_s}{ M_\chi M_N}\,.
\eeq
As expected, the heavier $\chi$  is, the smaller the mixing with the SM active neutrino is. Furthermore, its effect on the active neutrino mass is also less.
In many aspects, it behaves very much like a sterile neutrino, although it originates from a hidden sector.

It is important to note that $\Lag_5$ also gives rise to dimension-4 operators when only one of the
scalar fields picks up a VEV. An example is $\frac{v_s}{M_N}f_{R}\ovl{\chi_{R}} \ell_L H$, which is not present in the original Lagrangian. This can lead to invisible decay modes for the Higgs boson if $\chi$ is sufficiently light. However, the effective coupling is expected to be $\ord (10^{-10})$, and the decay cannot be detected in the near future. Similar terms can be read off from Eq.(\ref{eq:Lag5}), and they all have seesaw suppressed couplings.

Generalizing to the three active neutrino case is straightforward. For simplicity, we set $f_L=f_R=f$, and the neutral fermion mass matrix is now a $5\times 5 $ matrix since $\chi$ is now split into two Majorana
fermions denoted by $\chi_{1,2}$.
In the weak interaction basis, $\{ \nu_\alpha ( \alpha=e,\mu,\tau)\,, \chi_L,\chi_R^c\}$, and ignore the seesaw suppressed Majorana masses to $\chi_{R,L}$, this is given by
\beq
\label{eq:numass5}
\mathcal{M}_\nu\simeq \begin{pmatrix} \frac{y_{ee}v^2}{M_N}&\frac{y_{e\mu}v^2}{M_N}&\frac{y_{e\tau}v^2}{M_N}& \frac{y_{ef}vv_s}{M_N}&\frac{y_{e f}vv_s}{M_N}\\
\frac{y_{e\mu}v^2}{M_N}&\frac{y_{\mu\mu}v^2}{M_N}&\frac{y_{\mu\tau}v^2}{M_N}& \frac{y_{\mu f}vv_s}{M_N}&\frac{y_{\mu f}vv_s}{M_N}\\
\frac{y_{e\tau}v^2}{M_N}&\frac{y_{\mu\tau}v^2}{M_N}&\frac{y_{\tau\tau}v^2}{M_N}&\frac{y_{\tau f}vv_s}{M_N}&\frac{y_{\tau f}vv_s}{M_N}\\
\frac{y_{e}fvv_s}{M_N}&\frac{y_{\mu}fvv_s}{M_N}&\frac{y_{\tau}fvv_s}{M_N}&0& M_\chi \\
\frac{y_{ef}vv_s}{M_N}&\frac{y_{\mu f}vv_s}{M_N}&\frac{y_{\tau f}vv_s}{M_N}&  M_\chi&0\\
\end{pmatrix}\,,
\eeq
where we have restored the various Yukawa couplings of active neutrino $N_R$ couplings and $y_{\alpha\alpha^\prime}=  y_\alpha y_{\alpha^\prime}, y_{\alpha f}= y_\alpha f$.
The mass basis $\nu_i,i=1\cdots 5$
is related to the weak basis by a unitary transformation : $\nu_\alpha =\sum_i U_{\alpha i}\nu_i$. This diagonalizes Eq.(\ref{eq:numass5}), i.e. $M^{diag}_\nu =U^\dagger\mathcal{M}_\nu U$.
The weak charged current in the mass basis can be obtained from
\beq
\label{eq:CC}
\frac{ig}{\sqrt 2}\sum_{\alpha=e,\mu,\tau}\sum_{i=1}^5 U_{\alpha i}\overline{e}_\alpha \gm_{\mu L} \nu_i W^{\mu,-} + h.c.\,.
\eeq
Without going into the details of the numerical analysis of neutrino oscillation data, one can expect that $U_{\alpha4,5 }$  is approximately given by Eq.(\ref{eq:chinumix}).
Similarly, with the help of the unitarity of $U$, the neutral current involving neutrinos in the mass basis can be deduced as
\beq
\label{eq:NC}
\frac{i g}{2\cos \theta_w}\left[ \sum_{i=1}^{5}\bar{\nu_i}\gm_{\mu L} \nu_i -\left(\sum_{i,j=1}^{5}(U^\dagger)_{j\chi_L }U_{\chi_L i}\bar{\nu}_j \gm_{\mu L} \nu_i+\sum_{i,j=1}^{5}(U^\dagger)_{j\chi_R^c }U_{\chi_R^c i}\bar{\nu}_j \gm_{\mu L} \nu_i\right)
\right] Z^\mu.
\eeq
For $i,j\neq 4,5$ there is a small off-diagonal coupling that can be neglected. For the diagonal term the coupling strength is essentially that of the SM since the second term is negligible.

It is clear that since $\chi$ can mix with the active neutrino with a small mixing angle, it behaves very much like a sterile neutrino.   We emphasize that although $\chi$ and the commonly studied sterile neutrino  have similar  charged and neutral current interactions, their origins are very different.  There are numerous models for sterile neutrinos.  It is useful to compare our case with a complete model of the sterile neutrino to bring out the differences. A well-known example is the Neutrino Minimal Standard Model ($\nu$MSM) \cite{NuMSM}, which is a low scale seesaw model, i.e.,  the lepton number breaking scale is below the Fermi scale. Three sterile neutrinos  correspond to the three $N_R$'s,  two of which are in the $\mgev$ range, and the third is the $\mkev$ range. The last one is identified as WDM. It is the only state in this mass range. On the other hand, $\chi$ consists of two nearly degenerate states.
If the splitting is  $ > 100 \mkev$, optimistically, this can be seen in near future experiments(see Sec.\ref{sec:4}) and thus offers a distinction from the sterile neutrino scenario. Otherwise, $\chi$ will be a pseudo-Dirac fermion. In this case, it will be more difficult to tell the two scenarios apart. It may be necessary to examine other signals.

\section{ Stability of the Shadow fermion}
 \label{sec:3}
It is clear from the previous discussions that none of the hidden sector fields can be stable after SSB. For $\chi$ to play the role of WDM, we assume that $\chi$ is the lightest of the hidden particles. Denoting the mass eigenstates by
$\chi_\pm$ and requiring that they have masses in the $\ord (10\mkev)$ range, the only available decays are $\chi_\pm\ra 3\nu$ and $\chi_\pm \ra \nu\gm$. The width of the invisible decays is given by
\beq
\label{eq:chiinv}
\Gm(\chi_\pm\ra 3\nu)=\frac{G_F^2 M_{\chi_\pm}^5}{96 \pi^3}\sum_{i=1}^{3}\left|U_{\chi i}\right|^2
\eeq
as they are Majorana states.

The radiative decay is a 1-loop effect. Unlike most loop effects, the model dependence can be reduced to only the mixing angle by calculating the width in the $U$-gauge. The width is given by\footnote{ This is in agreement with the result of \cite{PW}. The calculation there was done in the
Feynman gauge and is valid for the sterile neutrino in the Type-I seesaw model. Our U-gauge calculation shows that this result holds for any SM singlet fermion that mixes with the active neutrinos, independent of how the individual masses are obtained.}
\beq
\label{eq:chinugm}
\Gm(\chi_\pm\ra \gm\nu)= \frac{9\alpha G_F^{2}}{256\pi^4}M_{\chi_\pm}^5 \sum_{\alpha=e,\mu,\tau}\sum_{i=1,2,3}|U_{4\alpha}|^2|U_{\alpha i}|^2\,.
\eeq
The invisible decays will be faster than the radiative mode. For $M_\chi =10\, \mkev$ and a lifetime longer than that of the Universe, we get the constraint
\beq
\label{eq:mixinglimit}
\sum_{i=1}^3 |U_{\chi i}|^2 < 1.8 \times 10^{-2}.
\eeq
This is to be compared with the expectation of $\sim 10^{-4}$ given by Eq.(\ref{eq:chinumix}) for
$v_s=1\mtev$. This validates $\chi_-$ as a WDM candidate. This decay will provide a monochromatic X-ray line for each of the $\chi$'s if the splitting is larger than the experimental resolution but still small
enough to be in the $< 10 \mkev$ range.

Next, we consider the scenario that $\chi$ splits into two Majorana neutrinos, one with mass $ 10\, \mkev$, and the other $100\, \mkev$. The lifetime of the heavier one is estimated to be
 $\sim 3.3 \times 10^{5}$ years with the same mixing as in Eq.(\ref{eq:mixinglimit}). This case will not affect the cosmic microwave background measurements and thus can also be a viable cosmological scenario.
 It may have later time cosmological implications that are beyond the scope of this investigation.

\section{Implications for low energy neutrino physics}
 \label{sec:4}
\subsection{$\beta$ decay spectrum}

It is well known that a detailed study of the $\beta$ decay spectrum of nuclei can reveal the
existence of one or more heavy neutrinos. This has been studied in the context of Kaluza-Klein extra-dimensional models \cite{McLN} where there can be many such neutrinos. More recently, a
detailed study has been conducted for tritium decays \cite{Kurie}. The KATRIN experiment \cite{KATRIN} can also be used to look for neutral leptons of mass lower than 18 keV.

\begin{figure}[htb!]
 \centering
 \includegraphics[width=0.3\textwidth]{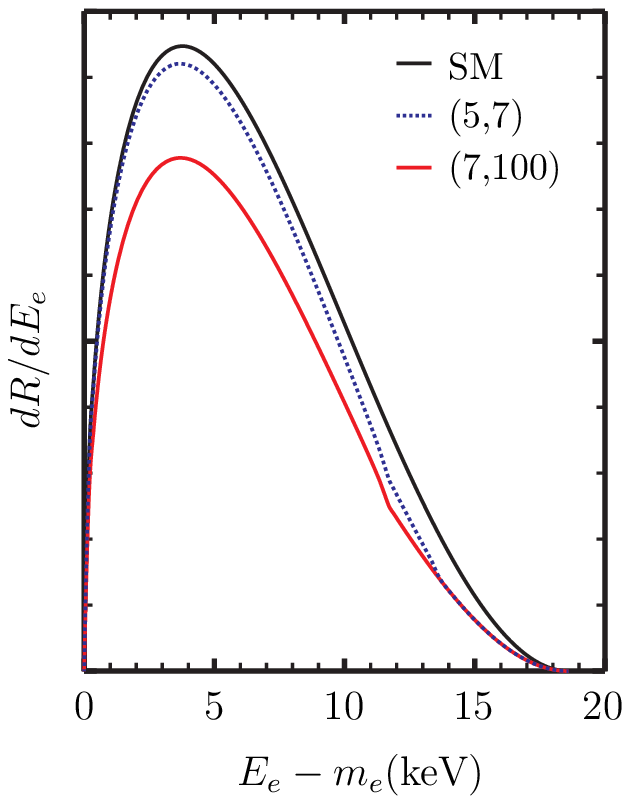} \quad
 \includegraphics[width=0.34\textwidth]{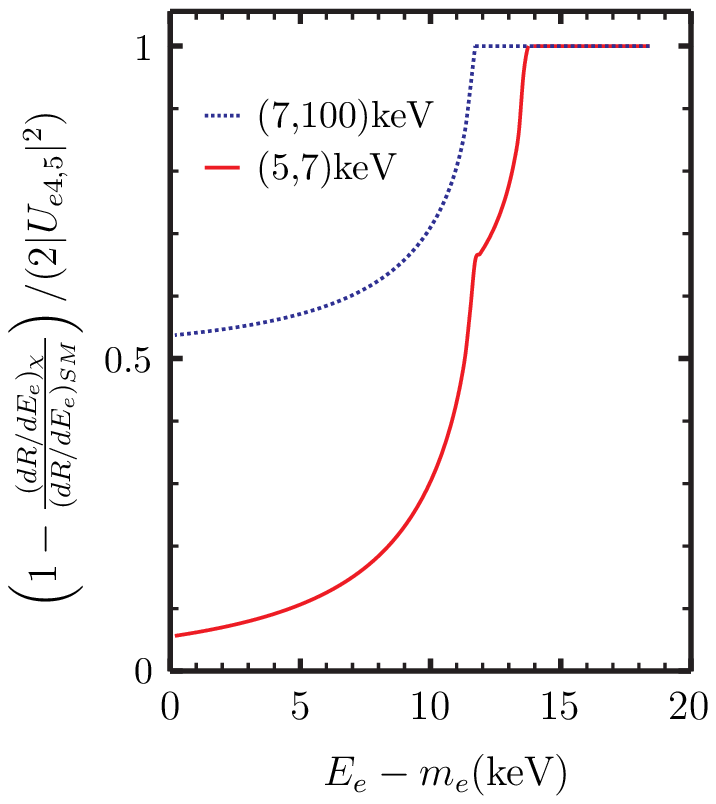}
\caption{ (a)   $dR/d E_e$, in arbitrary units, vs  $E_e-m_e$ (in keV). The black line
 denotes the SM curve. The dotted blue line is for $(M_1,M_2)= (7,5)\, \mkev$, and
 the red one is for $(M_1,M_2)= (7,100)\, \mkev$.
 The mixings are $U_{e4}=U_{e5}=0.4$.
(b) $1-\frac{(dR/d E_e)_\chi}{(dR/d E_e)_{\mathrm{SM}}}$  in units of $2|U_{e4,5}|^2$ with $(M_1,M_2)= (7,5)$ and $(7,100)\, \mkev$. }
\label{fig:kurie}
 \end{figure}

For a nuclear $\beta$ decay with mass $\gg Q,E_e,m_{\nu_\alpha}$, the  differential decay rate as a function of electron energy $E_e$ is given by the leading approximation
\beq
\label{eq:Kplot}
\begin{split}
\frac{d R}{d E_e}&= K_\beta E_e(Q+m_e-E_e)(E_e^2-m_e^2)^\half\\
&\left\{\left| U_{e5} \right|^2\left[(Q+m_e-E_e)^2-M_{2}^2\right]^\half +
\left|U_{e4}\right|^2\left[(Q+m_e-E_e)^2-M_{1}^2\right]^\half \right. \\
&\left. +\sum_{i=1}^3 \left|U_{ei}\right|^2\left[(Q+m_e-E_e)^2-m_{i}^2\right]^\half\right\}.
\end{split}
\eeq
where $K_\beta$ includes the nuclear matrix element, the Fermi function, and $G_F$.  Note that $Q$ is the $Q$-value of $18.59$ keV for tritium. We have also separated the heavy Majorana fermions $\chi_2,\chi_1$ with masses $M_2, M_1$ from the
physical active neutrinos, described by the last term. The spectrum consists of three branches if the energy resolution is smaller than $M_2-M_1$. The first one will cut off at
\beq
E_e=Q+m_e-M_2\,,
\eeq
and give a kink at that point. The second kink appears at
\beq
E_e=Q+m_e-M_1\,.
\eeq
 To illustrate, we display the differential decay rates in Fig.\ref{fig:kurie}(a) with an unrealistic sizable mixing $|U_{e4,5}|=0.4$ for two representative sets of $M_1,M_2$.

In Fig.\ref{fig:kurie}(b), we display the ratio of the same Kurie plot to that of the SM derivation from the unit for two representative sets of $M_1,M_2$.

If the energy resolution is not sufficient to resolve the two masses, then one smeared kink
will appear in the spectrum before it cuts off at $E_e\simeq 18.59 \mkev$ and gives the usual result of
\beq
\label{eq:mnue}
m_{\nu_e}= \sqrt{\sum_{i=1}^3 \left|U_{ei}\right|^2 m_i^2}\,.
\eeq
To be able to observe a kink-like structure in the spectrum, the energy resolution will have to
be approximately 300 eV in this energy range, and a dedicate experiment has been proposed \cite{TRISTAN}. A limit on the mixing of $|U_{e4,5}|^2 \lesssim 10^{-7}$ can be set if no kinks are found.

\subsection{$0\nu\beta\beta$ decays of nuclei}

For Majorana neutral fermions lighter than 100 keV that mix with the SM active neutrinos, the usual neutrino exchange mechanism for $0\nu\beta\beta$ decays of nuclei is still applicable. The effective Majorana mass for $\nu_e$ denoted by $m_{ee}$
is given by
\beq
\label{eq:mee}
 m_{ee}=\left|U_{e1}^2 m_1 + U_{e2}^2 e^{2i\alpha_2}m_2+U_{e3}^2 e^{2i\alpha_3} m_3
 +U_{e4}^2e^{2i\alpha_4} M_1 +U_{e5}^2 e^{2i\alpha_5} M_2 \right|\,,
 \eeq
 where $\alpha_j, j=2\dots 5$ are the Majorana phases. There are now four such phases in addition to the Dirac phases in $U_{ei}$ which now total 3. It is instructive to compare the contributions of the SM active neutrinos versus the $\chi$'s. Without
 going into detail, one expects the active neutrinos to contribute $10^{-2}-10^{-3}$ eV
 to $m_{ee}$. Using Eq.(\ref{eq:chinumix}) as a guide, they contribute $10^{-4}-10^{-2}$ eV, depending on the Yukawa, for $M_{1,2}\sim 10 \mkev$. Hence, the mixing of $\chi$ with the SM neutrinos will significantly change the expectations of $0\nu\beta\beta$ decays.

\subsection {Coherent low energy neutrino production of $\chi$}

We have argued that $\chi$ can be a warm dark matter candidate if its mass is in the keV range.
It is a prime candidate for production in low energy coherent neutrino-nucleus scattering (CE$\nu$NS), which has been observed recently in the COHERENT experiment \cite{COHERENT} at a spallation neutron source(SNS).
Such experiments can also be carried out at high powered reactors. The principal experimental challenge is to detect very low nuclear recoils. Cryogenic bolometers harbor the promise to detect sub-100 eV recoils \cite{Ricochet}.

The crucial physics requirement for CE$\nu$NS is a small momentum transfer to the nucleus.
It has to be smaller than the inverse radius of the nucleus to maintain coherence.
The scattering process must also not alter the quantum state of the nucleus. Nuclear excitations must not be triggered; otherwise, the process will break the coherence of nucleons that are scattered.
We studied the process
\beq
\label{eq:chico}
\nu + \mathcal{N} \ra \chi + \mathcal{N}\,,
\eeq
where $\mathcal{N}$ denotes the nucleus. We are interested in how Eq.(\ref{eq:chico}) can be used to limit the parameter space of the model.

\begin{figure}[h!]
\centering
\includegraphics[width=0.8\textwidth]{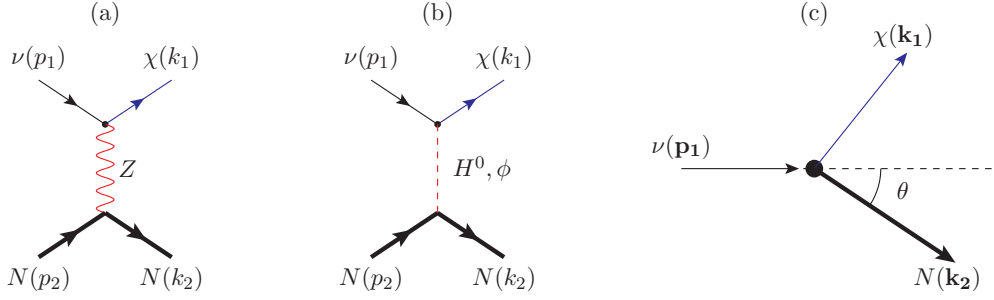} 
\caption{(a) and (b): The Feynman diagrams for $\chi$ production in the coherent neutrino-nucleus scattering.
The fixed target kinematics is depicted in (c). }
\label{fig:chiN}
\end{figure}

The two main fundamental processes are due to
$Z$ and $H$ exchanges, as shown in Fig.\ref{fig:chiN}.
The one-photon exchange process is very small in our model and can be ignored.
The two main processes have the same kinematics but give different differential cross-sections
$\frac{d\sigma}{dT}$ where $T$ is the recoil energy of the nucleus.
The $Z$ exchange process is similar to the SM coherent scattering and thus is suppressed by the $\nu-\chi$ mixing. It is well known that it is sensitive to the weak charge of the nucleus
$Q_w= N-Z(1-4\sin^2 \theta_W)$, and $N(Z)$ represents the number of neutrons(protons)in the nucleus.
Due to the accidental cancelation of the proton weak charge, this cross-section is expected to scale as $N^2$. Moreover, we can view Fig.\ref{fig:chiN}(a) as an additional branch to the SM process much in the same way it adds to the $\beta$ decay spectrum studied earlier.

On the other hand, Fig.\ref{fig:chiN}(b) has no SM equivalence since it
requires the existence of a righthanded singlet fermion such as sterile neutrino. Here $\chi$ plays that role, although it is not a sterile neutrino per se. Furthermore, Fig.\ref{fig:chiN}(b) is sensitive to
Higgs nucleon coupling, $ g_{Hnn}= \frac{g M_N}{2 M_W}\eta$, where $\eta\simeq 0.3$ \cite{HHG}, and $M_N$ is the nucleon mass.
This coupling is both experimentally and theoretically interesting.
 Roughly speaking, we can take $M\approx A M_N (A=N+Z)$ by neglecting the small neutron-proton mass difference.

The two important kinematic quantities are the scattering angle $\theta$ (see Fig.(\ref{fig:chiN} c)), and the recoil energy $T$. In terms of $T$, the scattering angle is
\beq
\label{eq: theta}
\cos \theta = \frac{E_\nu T + MT +\half m_\chi^2}{E_\nu \sqrt{T^2+2MT}}\,,
\eeq
where  $E_\nu$ is the incoming neutrino energy.
From this, we obtain the maximum $T_{+}$ and  minimum $T_{-}$ allowed recoil energies
\beq
\label{eq:Tpm}
T_{\pm}= \frac{ME_\nu^2 -\half m_\chi^2(M+E_\nu) \pm E_\nu \left[M^2E_\nu^2 -m_\chi^2 M(M+E_\nu)+\fth m_\chi^4\right]^\half}{M(M+2E_\nu)}\,.
\eeq

We recover the SM values by taking $m_\chi=0$. Hence $T_{+}^{\mathrm{SM}}= \frac{2 E_\nu^2}{(M+2E_\nu)}$ and $T_{-}^{\mathrm{SM}}=0$.
The differential cross-section can be written as
\beq
\label{eq:dsigg}
\frac{d\sigma^a}{d T}= \frac{1}{32\pi}\frac{1}{ME_\nu^2}\left[\half \sum_{\mathrm{spins}}|\mathcal{M}|^2\right]\,,
\eeq
where $\mathcal{M}$ is the invariant amplitude. The cross sections are the same
for $\bar{\nu}$.
For the $Z$ exchange process, we have
\beq
\label{eq:Zx}
\frac{d\sigma^{(Z)}}{d T}= \frac{G_F^2 Q_W^2 |U_{\ell 4}|^2}{4\pi}M \left[1-\frac{MT}{2E_\nu^2}-\frac{T}{E_\nu}+\frac{T^2}{2E_\nu^2} -\frac{m_\chi^2}{4E_\nu^2}\left(\frac{2E_\nu}{M}-\frac{T}{M}+1\right)\right]F_Z^2(q^2)\,,
\eeq
where $F$ is the nuclear form factor for the specific nucleus used in the detector, and $q^2=-2MT$ is the momentum transfer squared. For $m_\chi=0$, it reduces to the well-known SM result \cite{Freedman}.
Here $\ell$ is the flavor of the incoming neutrino.  For reactor neutrinos $\ell=e$, and $\ell= \mu, e$ for a spallation neutron source .
Similarly, the cross-section due to scalar exchanges is
\beq
\label{eq:Hx}
\frac{d\sigma^{(H)}}{d T}= \frac{y_\chi^2 g_{HNN}^2}{4\pi}\ M\cos^2{\alpha}\left(\frac{1}{M_H^2}+\frac{v}{v_s}\frac{\tan{\alpha}}{ m_\phi^2}\right)^2\left(1+\frac{T}{2M}\right)\left( \frac{M T}{E_\nu^2}+\frac{m_\chi^2}{2E_\nu^2}\right)F_{H}^2(q^2)\,,
\eeq
where $y_\chi$ parameterizes the new dimension-4 $H-\nu-\chi$ coupling (see Fig.(\ref{fig:seesaw})), and $\alpha$ is the mixing angle of the Higgs and $\phi$.
We have included $\phi$ exchange, although it is suppressed by
the ratio of VEV's and the mixing $\alpha$, as seen above. In the range of   $m_\phi \sim \mathcal{O}(1) \mgev$ ,
it can be comparable to the Higgs exchange effect. The Higgs coupling to nucleus  $g_{HNN}$  is an
unknown quantity. However, we take it to be  $g_{Hnn}$  with the substitution $m_n \ra M$.
Admittedly this is a gross attempt to capture the nucleon coherent effect. Furthermore, in general, the form factor $F_H$ is different from $F_Z$ in Eq.(\ref{eq:Zx}).

For the signal, we have to integrate over the appropriate neutrino spectrum. In general, the differential number of events per unit time
is given by
\beq
\label{eq:evrate}
 \frac{d N^{(a)}}{d T}=n_i \int^{E_{\nu \mathrm{max}}}_{E_{\nu \mathrm{min}}}d E_\nu\, \phi(E_\nu)\, \frac{d\sigma^{(a)} }{dT} (T,E_\nu)\,,
\eeq
where $n_i$ is the number of target nuclei in the detector, $\phi(E_\nu)$ is the flux of the incoming neutrinos, and $E_{\nu \mathrm{max}}$ is the maximum source
neutrino energy. The differential cross-sections for $a=\mathrm{SM}, Z, H$  can be read off from Eqs.({\ref{eq:Zx}) and (\ref{eq:Hx}).
The minimum required neutrino energy for a specific recoil $T$ is given by
\beq
E_{\nu \mathrm{min}} = \frac{ M T+ m_\chi^2 /2}{ \sqrt{2 M T +T^2}-T } \simeq \sqrt{M T /2}\,.
\eeq
 For a target with atomic number $A$ and recoil energy of $T=1 \mkev$, the minimal required neutrino energy is $\sim 7 \times \sqrt{A/100}\, \mmev$. For a neutrino source with energy $E_\nu$, and $m_\chi, E_\nu \ll M$, the maximal recoil energy is about
 $T_+\sim 20 \times(E_\nu/\mmev)^2\times(100/A)$ eV.
 Therefore, the maximal recoil energy in the $\nu+\mathcal{N}\ra \chi+\mathcal{N}$ process is about ${\cal O}(1)$ and ${\cal O}(40)\,\mkev$ for neutrinos from a nuclear power plant and a spallation neutron source, respectively.

The COHERENT experiment \cite{COHERENT} utilizes neutrinos from a spallation source. There are three flavors of incoming neutrinos from $\pi^+$ decays almost at rest into
$\mu^+ + \nu_\mu$ and the muon subsequently decays into $e^+\, \ovl{\nu_\mu}\, \nu_e$. The $\nu_\mu$ from the first decay gives a monochromatic flux. The muon decays are usually taken to be almost at rest. However, it is easy to take into account the energy of the muon which is given by $E_\mu=\frac{m_\pi^2 + m_\mu^2}{2m_\pi}=109.78 \mmev$.
The fluxes are calculated to be
\begin{subequations}
\beqa
\label{eq:flnumu}
\phi_{\nu_{\mu}}(E_\nu) &=& \phi_0\, \delta\left( E_\nu- \frac{m_\pi^2-m_\mu^2}{2m_\pi} \right)\,, \\
\label{eq:flnue}
\phi_{\nu_e}(E_\nu)     &=& \phi_0 \frac{192}{m_\mu}\left(\frac{E_\nu E_\mu}{m_\mu^2}\right)^2\left[\frac{m_\mu}{2E_\mu} -\frac{E_\nu}{m_\mu}\left(1+\frac{\mathbf{p}_\mu^2}{3E_\mu^2}\right)\right]\, ,\\
\label{eq:flnubar}
\phi_{\bar{\nu}_\mu}(E_\nu) &=& \phi_0\frac{64}{m_\mu}\left(\frac{E_\nu^2}{m_\mu^2}\right)\left[\frac{3}{4} -\frac{E_\nu E_\mu}{m_\mu^2}\left(1+\frac{\mathbf{p}_\mu^2}{3E_\mu^2}\right)\right]\,,
\eeqa
\end{subequations}
where  $\mathbf{p}_\mu$ is the muon 3-momentum and numerically $|\mathbf{p}|=29.79 \mmev$, and also $E_{\nu \mathrm{max}}=\half E_\mu$.
Note that $\phi_0$ is a normalization factor which depends on factors such as number of protons on target and the number of pions produced per incident proton. Specific to the COHERENT experiment, $\phi_0= \frac{r N_{\mathrm{POT}}}{4\pi L^2}$. The number of protons on target $N_{\mathrm{POT}}=1.76\times 10^{23}$, $r=0.08$ is the number of neutrinos per flavor produced for each proton on target, and $L=19.3$ m is the
distance between the source and the CsI detector. Also, the number of target nuclei in Eq.(\ref{eq:evrate}) is given by $n_{\mathrm{CsI}}= \frac{N_A M_{\mathrm{det}}}{M_{\mathrm{CsI}}}$, where $N_A$ is the Avogadro number,
$M_{\mathrm{det}}= 14.6$ kg is the detector mass, and $M_\mathrm{CsI}=259.8$ is the molar mass of CsI.
For the total signal, one multiplies Eq.(\ref{eq:evrate}) by the lifetime of the experiment. An acceptance factor that depends on $T$ is omitted but can be easily included.

In order to get a setup independent prediction, we can focus on the ratios of the signal with $\chi$ to the expected SM one.
 We first focus on the Z-mediated CE$\nu$NS.
For our model, the total differential rate consists of the ones from SM neutrinos and those from $\chi_\pm$,
  \beqa
  \frac{d N_\chi}{d T} &=& \frac{d N_{SM} }{d T}\times(1-|U_4|^2-|U_5|^2)+|U_4|^2\frac{d N^{(Z)}(M_1)}{d T}+|U_5|^2\frac{d N^{(Z)}(M_2)}{d T}\nonumber\\
  &=&\frac{d N_{SM}}{d T} +|U_4|^2\left(\frac{d N^{(Z)}(M_1)}{d T}-\frac{d N_{SM}}{d T}\right) +|U_5|^2\left(\frac{d N^{(Z)}(M_2)}{d T}-\frac{d N_{SM}}{d T}\right)\,.
  \eeqa
Note that here we use a notation that is slightly different from Eq.(\ref{eq:evrate}) with the $\chi-\nu$ mixing squared factored out, namely, $\frac{d N^{SM}}{d T} =\frac{d N^{(Z)}(m_\chi=0)}{d T}$.
For notational simplicity, we have also dropped the flavor content of the incoming neutrinos in the mixing elements,
the other notation is obvious.
 Therefore, the ratio of the total differential rate to the SM one deviates from 1 by an amount of
 \beq
 1- {\frac{d N_\chi}{d T}\over  \frac{d N_{SM}}{d T} }=
 |U_4|^2\left(1- \frac{d N^{(Z)}(M_1)}{d T}\bigg/ \frac{d N^{(Z)}(0)}{d T}\right) +|U_5|^2\left(1-\frac{d N^{(Z)}(M_2)}{d T}\bigg/ \frac{d N^{(Z)}(0)}{d T}\right)\,.
 \eeq

For simplicity, we assume $|U_4|^2=|U_5|^2= |U_\chi|^2$. Then, at fixed $T$,
\beq
\left(1- \frac{d N_\chi}{d T} \bigg/  \frac{d N_{SM}}{d T} \right)|U_\chi|^{-2}
=\left[2- \frac{d N^{(Z)}(M_1)}{d T}\bigg/\frac{d N^{(Z)}(0)}{d T}- \frac{d N^{(Z)}(M_2)}{d T}\bigg/\frac{d N^{(Z)}(0)}{d T}\right]\,,
\eeq
which is displayed in Fig.\ref{fig:nu_N_exp}(a) and (c) for SNS\footnote{ We have included the acceptance function of the COHERENT experiment\cite{ACCEPT} for the estimation. } and nuclear power plant neutrino sources, respectively. The reactor neutrino flux in \cite{KS_taiwan} is adopted.
However, note that there is no SM counterpart for the scalar-mediated coherent scattering, and the mixings are different from the Z-mediated ones.
Therefore, the scalar-mediated part is separately compared to the Z-mediated SM one,
\beq
\left(  \frac{d N_H}{d T} \bigg/  \frac{d N_{SM}}{d T} \right)|U_H|^{-2}
= \left[ \frac{d N^{(H)}(M_1)}{d T}\bigg/\frac{d N^{(Z)}(0)}{d T}+ \frac{d N^{(H)}(M_2)}{d T}\bigg/\frac{d N^{(Z)}(0)}{d T}\right]\,.
\eeq
Again, we assume the couplings of the two shadow fermions are the same, and the quantity $|U_H|^2$ is defined as
\beq
|U_H|^2\equiv { y_\chi^2 g_{HNN}^2  \over M_H^4 G_F^2 Q_W^2} P_\chi^2 \simeq 7\times 10^{-5} \left(\frac{A}{N}\right)^2 y_\chi^2  P_\chi^2\,,
\eeq
where
\beq
P_\chi =c_\alpha +s_\alpha \frac{v}{v_s}\frac{M_H^2}{M_\phi^2}
\eeq
is of order unit as long as $s_\alpha \lesssim 10^{-3}$ and $M_\phi\sim {\cal O}(\mbox{GeV})$. If taking $P_\chi^2\sim 1$, then $|U_H|^2 \sim 10^{-4}\times y_\chi^2$.
 On the other hand, if $s_\alpha>10^{-3}$,  the process will be dominated by the light $\phi$ and
$P_\chi^2 \gg 1$. These contributions are displayed in Figs.\ref{fig:nu_N_exp}(b) and (d) for SNS and nuclear power plant neutrino sources, respectively.

Note that the scalar-mediated CE$\nu$NS is not very sensitive to $M_{1,2}$.
Also, observe the jumps at around $T\sim 13$keV in the SNS CE$\nu$NS. They are due to a sharp $T_{max}$ cutoff from the monochromatic muon neutrino line, Eq.(\ref{eq:flnumu}).
 \begin{figure}[hbt!]
 \centering
 \subfigure[]{\includegraphics[width=0.4\textwidth]{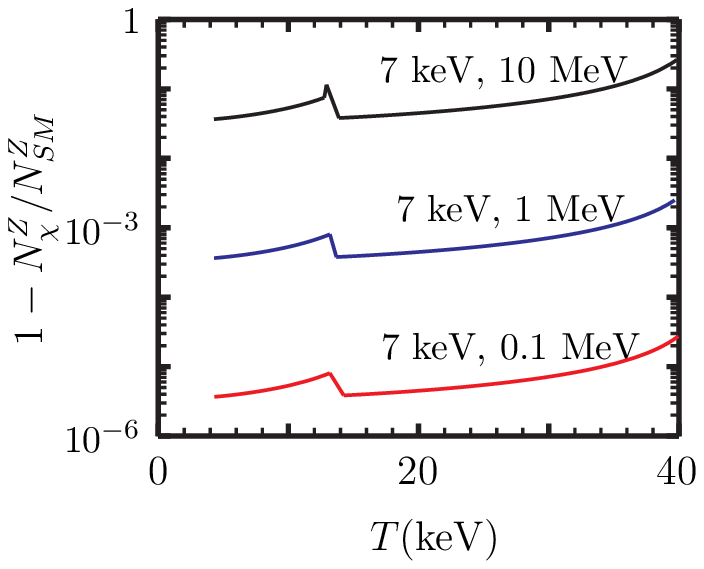}} \quad
  \subfigure[]{\includegraphics[width=0.4\textwidth]{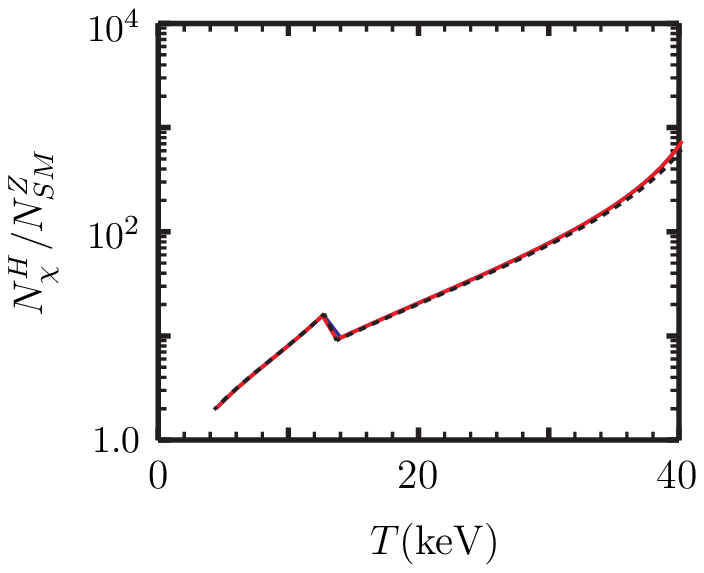}}\\
   \subfigure[]{\includegraphics[width=0.4\textwidth]{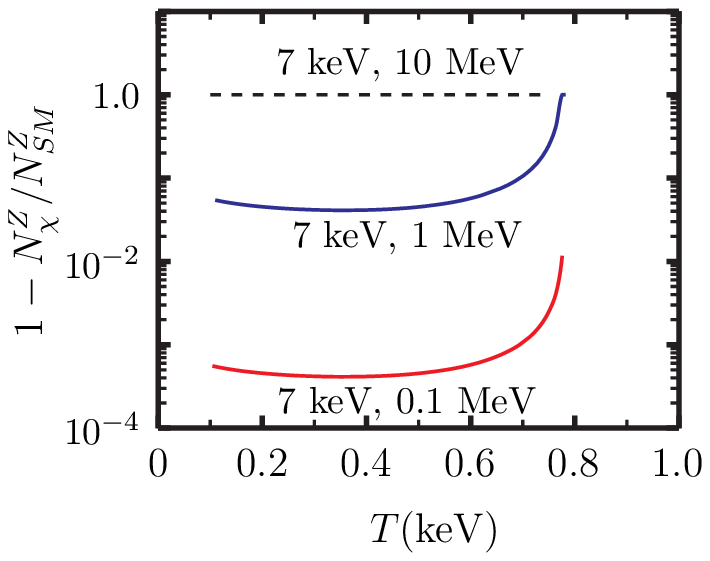}} \quad
 \subfigure[]{\includegraphics[width=0.4\textwidth]{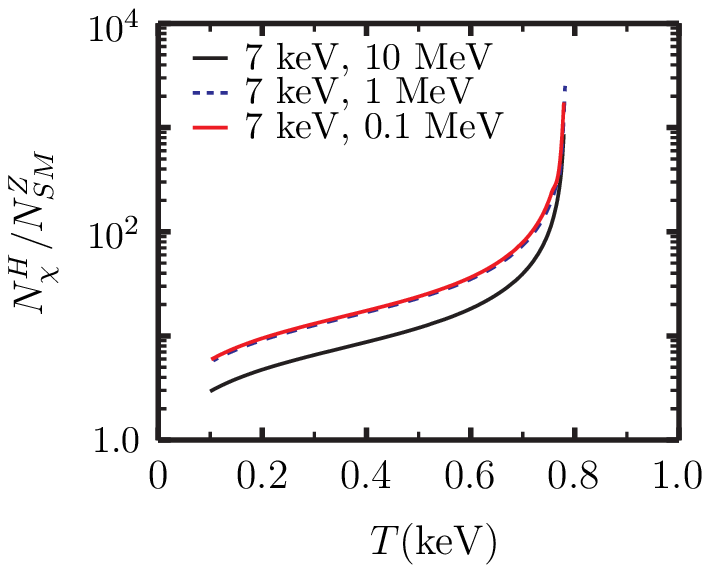}}
\caption{(a) The deviation from the SM CE$\nu$NS at the SNS neutrino source from $Z-$mediated scattering on the CsI target.
Here we consider the simplified scenario where  $|U_{4,5}|^2 =|U_\chi|^2$.
(b) Same as (a) but mediated by the scalar exchange. The deviations are in units of $|U_\chi|^2$ and $|U_H|^2$ for sub-diagrams (a) and (b), respectively. In other words, we set $|U_\chi|^2=1$ and $|U_H|^2=1$ for the plots.
(c, d): Same as (a) and (b) but with a nuclear power plant neutrino source.}
\label{fig:nu_N_exp}
 \end{figure}

Assuming that the Z-mediated process dominates, from Fig.\ref{fig:nu_N_exp}, one sees that the SNS experiments are more sensitive to  the $M_{1,2}$ splitting. For the small splitting cases, such as  $\{M_1,M_2\}=\{7, 100\} \mkev$, this will be extremely challenging for the SNS experiments. On the other hand, the signal is about a thousand times bigger at the reactor experiments but requires a very low threshold for recoil energy, i.e., $T< 1 \mkev$.
However, this is beyond current capabilities for most proposed experiments. The successful development of
cryogenic detectors such as the proposed \cite{nu-cleus} experiment may bring these measurements
to reality. As compared to Kurie plot experiments,
neutrino-nucleus coherent scatterings cannot probe $\chi$ splittings less than $100 \mkev$ in the foreseeable future. On the positive side, they are sensitive to low energy scalars that can mix with the Higgs boson.
These fields are popular in Higgs portal constructions but are notoriously difficult to get an experimental
handle on.

\section{Electroweak Precision Tests}
 \label{sec:5}
\subsection{Effects of kinetic mixing of $U(1)_Y$ and $U(1)_s$}
  The second portal of $X_\mu$ can be studied independently of the other two. First, we note that the kinetic terms, including the mixing, can be recast into canonical form through a $GL(2)$ transformation. Explicitly, this is given by
\beq
\left(\begin{array}{c} X\\ B \end{array}\right)
= \left(\begin{array}{cc} c_\epsilon & 0 \\ - s_\epsilon & 1 \end{array}\right)
\left(\begin{array}{c} X'\\ B' \end{array}\right) \,,
\eeq
where
\beq\label{Eq:KinMix}
s_\epsilon = \frac{\epsilon}{\sqrt{1 - \epsilon^2}}\,,\;
c_\epsilon = \frac{1}{\sqrt{1 - \epsilon^2}} \,.
\eeq
After SSB, $X'$ and $B'$ will mix and result in a shift in the SM $Z$ mass. The physical neutral bosons consist of three states $\gm, Z, Z_s$. The transformation relating the weak and mass bases is given by
\beq
\left( \begin{array}{c} B'\\A_3\\ X' \end{array}\right)
=\left( \begin{array}{ccc} c_W & -s_W& 0\\
s_W & c_W & 0\\ 0&0&1 \end{array}\right)
\left( \begin{array}{ccc} 1&0&0\\ 0& c_\eta & -s_\eta\\
0& s_\eta & c_\eta \end{array}\right)
\left( \begin{array}{c} \gamma \\ Z \\ Z_s \end{array}\right) \,,
\eeq
where $s_W$ ($c_W$) denotes $\sin\theta_W$ ($\cos\theta_W$) and similarly for the rotation angle $\eta$. The first rotation is the standard one that gives rise to the SM $Z$, and the second one diagonalizes the mixing of the two $Z$ bosons. The extra mixing angle is given by
\beq
\label{eq:taneta}
\tan {2\eta} = \frac{2 s_W s_\eps}{c_W^2 (M_3/M_W)^2 + s_W^2 s_\epsilon^2 - 1} \,,
\eeq
where $M_3 \equiv g_sv_s$ and $\left\langle \phi \right\rangle = \frac{v_s}{ \sqrt{2}} $.
We use the shorthand notation $c_\eta(s_\eta) = \cos \eta (\sin{\eta})$ and $t_{2\eta}= \tan {2\eta}$. In general, $\epsilon$ is a free parameter; however, the success of the SM indicates that it has to be small.
Clearly, the photon will remain massless, and the $W$ bosons will be unchanged from the SM. Our notations follow that of \cite{CNW1} where details can be found.

The masses for the two massive neutral gauge bosons are readily found to be \cite{CNW1}:
\beqa
\label{eq:Zmasses}
M^2_{Z}&=& M_Z^{2\msm}
\left\{c_\eta^2 -s_{2\eta}s_W s_\eps + s_\eta^2 s_W^2 s_\eps^2\right\} +s_\eta^2 M_3^2\,, \non\\
M_{Z_s}^2 &=& M_Z^{2\msm}\left\{s_\eta^2 +s_{2\eta}s_W s_\eps +c_\eta^2 s_W^2 s_W^2 s_\eps^2\right\}+c_\eta^2 M_3^2 \,.
\eeqa
We see that the Z-boson has its mass shifted from the SM value of $M^{SM}_Z=\frac{M_W}{c_W}$, and receives a small contribution from the hidden sector. Similarly, the physical $Z_s$ mass comes mainly from the hidden sector with
a small contribution from the visible sector. We see later that if we want $\chi$ to be warm dark matter, it is more natural to have $M_{Z_s} \gg M_W$. With that in mind, Eqs.(\ref{eq:taneta}) and (\ref{eq:Zmasses}) become
\beqa
\label{eq:approx}
\eta &\sim& \frac{s_W s_\eps}{c_{W}^2}\frac{M_{W}^2}{M_3^2}\lesssim s_\eps\,, \\
M_Z^2 &\sim& \left(\frac{M_W}{c_W}\right)^2 \left(1- \frac{s_W^2 s_\eps^2}{c_W^2}\frac{M_W^2}{M_3^2}\right)\,.
\eeqa
The very precise measurement of the $\rho$ parameter gives a stringent limit on the mixing parameters. The shift $\delta \rho$
is
\beq
\label{eq:rhoc}
\delta \rho = \frac{s_W^2 s_\eps^2}{c_W^2}\frac{M_W^2}{M_3^2}
< 3.7 \times 10^{-4}\,.
\eeq
In turn, we get
\beq
\label{eq:M3}
M_3> 2.29\, |s_\eps|\, \mtev\,.
\eeq
 There are many electroweak precision tests (EWPT) that can set limits on $\eps,\eta$.
 In particular, the measurements at the Z-pole are independent of the mass of $Z_s$ but are sensitive to the modifications to the SM fermion-fermion-Z couplings. These couplings are flavor universal and explicitly given by
\beqa
\label{eq:zff}
Z^\mu \bar{f}f\; : \; i \gamma^\mu \frac{ g_2}{c_W}
\left[ \left( c_\eta  g^L_f   - s_\eta s_W s_\epsilon  Y_f^L  \right)
\PL\;\right.\non\\
\left. +\left( c_\eta g^R_f- s_\eta s_W s_\epsilon  Y_f^R   \right) \PR \right] \,,\\
\label{eq:zprimeff}
Z_s^\mu \bar{f}f\; : \;i \gamma^\mu \frac{ g_2}{c_W}
\left[ \left( -s_\eta  g^L_f   - c_\eta s_W s_\epsilon  Y_f^L  \right)
\PL\;\right.\non\\
 \left. +\left( -s_\eta g^R_f- c_\eta s_W s_\epsilon  Y_f^R   \right) \PR \right] \,,
\eeqa
where $g_{L,R}^f = T^3(f_{L,R}) - s_W^2 Q^f$ is the coupling of the SM $Z$ to fermions and
$\PL =\frac{1-\gm_5}{2},\PR=\frac{1+\gm_5}{2}$.
A comprehensive list of couplings used
to constrain $\eps$ can be found in \cite{CNW1}.

\subsection{Invisible $Z$ decays}

The two $U(1)$'s mixing will also modify the SM Z boson invisible decay width $\Gamma_{\mathrm{inv}}$. There are two changes : (i) the modification of the $Z-\nu-\nu$ couplings as given in Eq.(\ref{eq:zff}), and (ii) the opening of the new channel $Z\ra \ovl{\chi_{L}} \chi_{L},\ovl{\chi_{R}} \chi_{R}$ from the mixing with $Z_s$.
The new invisible width is
\beq
\label{eq:zchichi}
\Gamma(Z\ra \chi\chi)= \frac{c_\eps^2 s_\eta ^2 g_s^2}{12 \pi} M_Z\,,
\eeq
where we have neglected the masses of $\chi$. The experimental value of $\Gamma_{\mathrm{inv}}= 499 \pm 1.5 \mmev$ agrees well with three nearly massless active neutrinos. Thus
\beq
\label{eq:etalim}
s_\eta^2 \left[\frac{c_\eps^2 g_s^2}{3}- \frac{g_2^2}{8c_W^2}\right]\le 2.1\times 10^{-4}.
\eeq
Other precision measurement constraints such as muon $g-2$, atomic parity violation, and M\o ller scattering  are given in \cite{CNW1} and will not be repeated here.

\subsection{$Z\ra f\bar{f}\phi^0$ decays}
\label{sec:Zffphi}
As will be discussed in the next section,  for $\chi$ to be WDM, the new physical scalar $\phi^0$, mainly stemming from the real part of $\phi$, is expected to be light, $m_\phi\sim$ a few GeV, and long-lived, $\tau_\phi\sim 1$ sec. Through the Higgs portal, the SM Z boson can now have a tree-level 3-body decays  $Z\ra Z^* \phi^0 \ra \bar{f}f \phi^0$, where $f$ is the SM fermions.
 The mixing between $\phi^0$ and SM Higgs depends on $\kappa$  (see Eq.(\ref{eq:LT})) as well as other parameters in the scalar
potential. The details are not relevant here, and we will denote this resulting mixing by $\kappa_s$.
Thus, $\phi^0$ couples to the SM fields with strengths of the SM Higgs couplings times $\kappa_s$.
The decay branching ratio is calculated to be \cite{HHG,CNW}:
\beq
Br(Z\ra \phi^0 f\bar{f})= \kappa_s^2 \times {\cal F}(m_\phi/M_Z) \times Br(Z\ra f\bar{f})\  ,
\eeq
where
\begin{eqnarray}
{\cal F}(x)=\frac{G_F M_Z^2 }{24\sqrt{2} \pi^2 }
&& \left[ {3 x (20-8x^2+x^4)\over \sqrt{4-x^2}}\cos^{-1}\left( \frac{x}{2}(3-x^2)\right) - \right.\nonumber \\
&& \left. 3(4-6x^2+x^4)\ln x-\frac{1}{2}(1-x^2)(47-13x^2+2x^4) \right]\, .
\end{eqnarray}
Due to its long lifetime, $\phi^0$ will escape the detector, and the apparent signal will be $Z\ra\bar{f}f+\not\!\!E$.
The SM background will be $Z\ra \bar{f}f\bar{\nu}\nu$.
And $f=\mu, b$ are ideal options to search for such processes and probe the scalar mixing squared, $\kappa_s^2$, down to $10^{-7}-10^{-8}$ with  $10^{12}$ fiducial Z's \cite{CNW}. The smallest $\kappa_s$ that can be probed by this process is still roughly one order too big for $\phi^0$ to dilute the DM relic density, Eq.(\ref{eq:kapplim}). However, this search provides an interesting experimental cross-check for whether this model can accommodate the WDM in the way described in the next section.

\section{$\chi$ as warm dark matter}
 \label{sec:6}
The previous discussions are independent of whether and how $\chi$ can become WDM. Here
we  examine the parameter space that allows $\chi$ to become WDM. We intend to give a broad-brush
description of a possible scenario and will leave many interesting details for a future study.
 Most of the discussions given below do not depend on the fact that  physical $\chi_\pm$ are Majorana fermions.
If the splitting is small, then they will behave as one Dirac particle. If the splitting is large, then
only the lighter one $\chi_{-}$ will serve as DM.

For clarity, we take $\chi$ to be Dirac. The Majorana
case can be obtained by $\bar{\chi}\ra \chi^c$, and we use chiral projections where needed.
 We specifically
explore the hierarchy of scale $M_{Z_s} \gtrsim M_Z \gg M_\phi >M_\chi$. The small $Z_s-Z$ mixing will be denoted by $\eta$,
see Eq.(\ref{eq:taneta}). After SSB of $U(1)_s$, the physical degrees of freedom in the hidden sector are $Z_s,\chi ,\phi^0$, where $\phi^0$ is the physical scalar mainly constituted by $\Re \phi$. We also use the benchmark values $M_\chi = 10\, \mkev$ and $M_\phi = 2\, \mgev$ to focus our discussion.

The secluded sector and the SM have feeble interactions through the portals. If the portal connections are switched off, the two sectors will not establish thermal equilibrium, and cosmologically they evolve separately. If the portal interactions are small but not negligible, then the SM and the hidden sector can interact. In particular, $\chi$ can interact
with the SM fields via the $Z_s-Z$ mixing. The process $\chi \bar{\chi}\leftrightarrow f \bar{f}$, where $f$ is a SM fermion, can proceed via such a mixing.  The cross-section at temperature $T$ can easily be estimated to be
\beq
\label{eq: xxff}
\sigma v\sim \left(\frac{\eta g_s}{g_2}\right)^2 G_F^2 T^2 \equiv A_\chi^2 G_F^2 T^2\,,
\eeq
where $g_2$ is the SM $SU(2)_L$ gauge coupling. If this rate falls below the Hubble expansion rate, $\chi$ will freeze out.
 The freeze-out temperature $T_f$ can be estimated by
setting $n \langle \sigma v \rangle$ to be equal to the Hubble expansion rate, and $n$ is the number density of the freeze-out particle. Thus,
\beq
\label{eq:Tf}
\frac{3 \zeta(3)}{2\pi^2} A_\chi^2 G_F^2 T_f^5 =\left(\frac{8\pi^3}{90}\right)^\half \frac{T_f^2}{M_{pl}} \sqrt{g_*}\,,
\eeq
with $g_*$ being the effective number of degrees of freedom. It is given by
\beq
\label{eq:gstar}
g_*= \sum_{\mathrm{bosons}} g_i +\frac{7}{8} \sum_{\mathrm{fermions}} g_i\,,
\eeq
and $g_i$ is the number of spin states.

From Eq.(\ref{eq:Tf}), the freeze-out temperature of $\chi$ is controlled by the parameter $A_\chi$. From the
electroweak constraints, Eq.(\ref{eq:etalim}), $\eta g_s \lesssim \mathcal{O} (10^{-2})$. Hence, it is natural to take $A_\chi=0.01$, and
we get
\beq
\label{eq:tfben}
T_f=87.5 \left(\frac{g_*}{100}\right)^{\frac{1}{6}}\left(\frac{0.01}{A_\chi}\right)^{\frac{2}{3}} \mmev.
\eeq
For a keV $\chi$, which we are interested in, it is relativistic at the freeze-out. Such a situation will lead to overclosure of the Universe.

To see this, we note that the number density per entropy is
\beq
\label{eq:Y}
Y\equiv \frac{n_\chi}{s}\simeq \frac{135 \zeta(3)}{4\pi^4 g_*(T_f)}\,.
\eeq
Note that $Y$ is thermally conserved and that it gives the relic density of $\chi$ as
\beq
\label{eq:omchi}
\Omega_\chi = \frac{Y m_\chi s}{\rho_c}\simeq 2.5 \times \left(\frac{m_\chi}{\mkev}\right)\left(\frac{100}{g_*(T_f)}\right)\,,
\eeq
where we have used the present-day entropy density $s=2891.2 \mathrm{cm}^{-3}$, critical density
$\rho_c= 1.05371\times 10^{-5} h^2 \mgev \mathrm{cm}^{-3}$ and $h=0.678$\cite{PDG2018}.
Comparing with the observed dark matter relic abundance\cite{PDG2018} of
\beq
\label{eq:Omdm}
\Omega_{\mathrm{DM}}=0.258\pm 0.011\,,
\eeq
the estimate of Eq.(\ref{eq:omchi}) clearly overcloses the Universe unless $g_*$ is of order 1000.
In our model $g_*(T_f)\sim 31$ since $T_f$ is below $1\, \mgev$.
Hence, some mechanism is required to bring down the value of $\Omega_\chi$.
Moreover, Eq.(\ref{eq:omchi}) shows that the higher $T_f$ is, the less severe the overclosure
problem is since $g_*$ will be larger.

One way to get around this obstacle is to dilute the density by producing more entropy. The dilution can come from the decay of the scalar $\phi^0$ if it has a long enough lifetime, and decays into SM particles during the era that $\chi$ is freezing
out.
Note that, due to parity, $\phi^0\phi^0$ cannot annihilate into SM fermions via the $Z_s-Z$ mixing.
However, it can do so via mixing with the Higgs boson.
It must be relativistic when $\chi$ freezes out at $T_f$ so that it has no
Boltzmann suppression at decay. It is easier to make such arrangement than adjusting $T_f$ given in Eq.(\ref{eq:Tf}).
The ballpark estimation is as follows. The $\phi^0\phi^0 H $ coupling is $\sim\kappa v$,  Eq.(\ref{eq:LT}), and gives
\beq
\label{eq:phiphimumu}
\sigma(\phi^0\phi^0 \ra \bar{\mu}\mu) \simeq \left(\frac{\kappa v}{M_H^2}\frac{m_\mu}{M_W}\right)^2\, .
\eeq
The freeze-out takes place when temperature yields $n\sigma \lesssim \frac{T^2}{M_{pl}}$, or
\beq
T\sim \kappa^{-2}\times 10^{-9}\, \mgev\,.
\eeq
Therefore, with $\kappa < 10^{-5}$, it is ensured that $\phi^0$ decouples relativistically and before $\chi$
freezes out.
As $\phi^0$ is unstable and can decay into SM fermions via mixing with Higgs,
its decay will transfer the energy density into radiation while doing so. Using the sudden decay approximation that
all the $\phi^0$'s decay at $t\simeq \tau_\phi$ and reheat the Universe to a temperature of
$T_r$\cite{ST},
\beq
\label{eq:tr}
T_r\simeq 0.78 \left(g_*(T_r)\right)^{-\frac{1}{4}}\sqrt{\Gamma_\phi M_{pl}}=1.11\, \mmev \left(\frac{1\mathrm{sec}}{\tau_\phi}\right)^{\half}.
\eeq
In the above,  we have used $g_*=10.75$ and taken $\phi^0$ lifetime to be $\sim 1$ sec.
The reason for taking this value is the constraint impose by Big Bang Nucleosynthesis(BBN). In order not to
upset a successful BBN, $T_r$ should be larger than 1 MeV, and this sets $\tau_\phi \lesssim \mathcal{O}(1)$ second.
Next, we use energy conservation
\beq
\label{eq:enc}
m_\chi Y s(T_f)= \rho(T_r)=\frac{3}{4} s(T_r) T_r
\eeq
and obtain the dilution factor
\beq
\label{eq: dilu}
D\sim\frac{s(T_r)}{s(T_f)}= 280.4 \times \frac{g_*(T_r)^{\frac{1}{4}}}{g_*(T_f)}\, \left(\frac{m_\phi}{1\, \mgev}\right)\, \left(\frac{1\,\mathrm{sec}}{\tau_\phi}\right)^\half.
\eeq
Inserting this dilution factor into Eq.(\ref{eq:omchi}), the relic density of $\chi$ is
\beq
\label{eq: relichi}
\ovl{\Omega_\chi}= \frac{0.89}{g_*(T_r)^{\frac{1}{4}}}\left(\frac{1\,\mgev}{m_\phi}\right)
\left(\frac{\tau_\phi}{1\,\mathrm{sec.}}\right)^\half\,.
\eeq
Interestingly, the dependence on $g_*(T_f)$ cancels out with this entropy dilution mechanism.
If we take $T_r$ to be slightly higher than 1 MeV, then Eq.(\ref{eq:gstar}) gives
 $g_*(T_r)=15.25$. And with $ m_\phi= 2\, \mgev $, one obtains the right amount of entropy dilution.

Note that $\phi^0$ can decay into SM particles via the mixing with the SM Higgs.
 As discussed in  Sec.\ref{sec:Zffphi}, we parameterize the unknown $\phi^0$-Higgs mixing as $\kappa_s$.
The width of $\phi^0\ra f \bar{f}$ where $f$ is a SM fermion is given by
\beq
\label{eq:phiwidth}
\Gamma = \kappa_s^2\frac{N_c \alpha m_f^2}{8 M_W^2} m_\phi \beta_f^3\,,
\eeq
where $N_c$ is the color of $f$, $\beta_f = \sqrt{ 1-4 m_f^2/m_\phi^2 }$, and $\alpha$ is the fine structure constant.
 However, for $m_\phi \simeq 2\, \mgev$, the main decay is into a gluon pair. The rate is estimated to be\cite{HHG}
\beq
\Gamma_{gg} = \left(\frac{\alpha_s}{3\pi}\right)^2 \frac{m_\phi^2}{m_\mu^2 } { [6-2\beta_\pi^3-\beta_K^3]^2 \over \beta_\mu^3} \Gamma_{\mu^+\mu^-}\,.
\eeq

 Demanding that $\tau_\phi \simeq 1 $ sec. leads to
\beq
\label{eq:kapplim}
10^{-5} \gtrsim \kappa_s \gtrsim 1.3 \times 10^{-9},
\eeq
whereas the upper bound comes from previous considerations.

While we have identified the parameter space for $\chi$ as a WMD, we still have to ensure
that there are no large processes that can generate significant numbers of $\chi$ during or
after thermal freeze-out. One process in which $\chi$'s can be produced is $\phi^0\phi^0 \ra \chi \bar{\chi}$.
This is suppressed by large seesaw mass $M_N$,  see Eq.(\ref{eq:Lag5}), and can be neglected. Another source will be $\phi^0\ra
\chi \bar{\chi}$. If assuming $f_L=f_R=f$, the effective coupling is given by $y_{eff}^{\chi}\equiv\frac{f^2 v_s}{M_N}$. We will require this mode to be less than the SM, and this leads to a loose bound
\beq
\label{eq:yeffchi}
y_{eff}^{\chi} < \kappa_s \frac{m_\mu}{M_W}\,.
\eeq

 A third process of producing $\chi$ is via active $\nu$ and $\chi$ oscillations \cite{DW,DH}.
The $\chi$ production rate via this mechanism peaks at temperatures of $\sim 0.1-1$ GeV.
This deserves a detailed study which is beyond the scope of this paper. We note that if this mechanism saturates
the bound on dark matter relic density, it will give an upper bound on the mixing between the active neutrino and the lighter $\chi_-$, $|U_{i\chi_-}|^2 \lesssim  10^{-9}  (10\, \mkev/M_\chi)$
\cite{snudm}. However, the entropy dilution mechanism operates at a lower temperature $\sim 1$ MeV, Eq.(\ref{eq:tr}),  and will likely loosen the above constraint.
Moreover, our model has one crucial difference from the $\nu$MSM. Although we have so far only described phenomenology by treating $\chi$ as a Dirac fermion, it is, in fact, two Majorana fermions. If the splitting is small, they
become pseudo-Dirac. Otherwise, they are a pair of Majorana fermions with very different masses. This can be arranged as discussed in Sec.\ref{sec:2}.
One can always arrange the model parameters such that the lighter one, $\chi_-$, fulfills all constraints and becomes the DM.
On the other hand, if the mass splitting is sizable, the heavier one, $\chi_+$, quickly decays. Therefore, constraints for $\chi_-$ do not apply to $\chi_+$.  Our previous discussions of the low energy experiments will now be applied to $\chi_+$ and can be used to set bounds on the mixing with active neutrinos.

 To conclude this section, we note that $\phi^0$ in the mass range of $ 1 - 10\, \mgev$ is notoriously challenging for experiments to discover. A direct detection will be impossible at the LHC \cite{CMN}. However, the Higgs boson invisible decay can be searched for via $H\ra \phi^0 \phi^0$, and the pair of $\phi^0$'s will act as missing energy as noted before. The width is
\beq
\label{eq:h2phi}
\Gamma(H\ra 2\phi^0)=\frac{\kappa_s^2 v^2}{32 \pi M_H} \sqrt{1-\frac{4 m_\phi^2}{M_H^2}}\,.
\eeq
 ATLAS\cite{ATinv} and CMS \cite{CMSinv} place a limit on invisible branching ratio ${\cal B}_{H\ra inv}\lesssim 24-30\%$ at 95\%C.L., which gives the constraint $\kappa_s \lesssim 0.02$. Hence, rare Z decays will still be the best probe of light and long-lived scalars.

\section{Conclusions}
 \label{sec:7}
We have explored a simple model of the secluded sector that consists of only a Dirac fermion $\chi$ charged under a $U(1)_s$.
It has vector couplings to $U(1)_s$ and hence is not anomalous. Furthermore, it can acquire a bare mass term $M_\chi
\bar{\chi}\chi$. Although $M_\chi$ is a free parameter, it is of particular interest if it has a value
$\lesssim 1 \mmev$ that will make it a candidate for warm dark matter.

The secluded sector can be connected to the SM via three kinds of portals.  The first is the seesaw portal (SP). It consists of three SM singlet righthanded neutrinos $N_R$'s which allows us to implement type-I seesaw for active neutrino masses. The second portal is due to the kinetic mixing of $U(1)_s$ and the SM hypercharge $U(1)_Y$. This is the gauge portal (GP). Note that $U(1)_s$ symmetry is broken spontaneously at a scale $v_s$ below the seesaw scale by a SM singlet scalar $\phi$. We also take $v_s>v$. The gauge invariant term $\phi^\dagger \phi H^\dagger H$ provides the third portal. This is commonly known as the Higgs portal (HP). Moreover, the gauge invariant $\chi N_R \phi$ Yukawa couplings provide an indirect connection between $\chi$'s and the SM sector
 after $N_R$ is integrated out.  All three portals have been discussed independently as simplified models
for dark matter. Taking all three together reveals features that are not present in the simplified 1-portal models.

SP not only provides the  seesaw mechanism for active neutrinos, but it also splits $\chi$ into two
Majorana fermions.  They can form  Dirac mass terms with the active neutrinos, and act very much like sterile neutrinos, although they are not.  These singlet fermions originate from the hidden sector. They add to the structure
of the neutrino mass matrix and contribute to $0\nu\beta\beta$ decays. They make their presence
felt in the Kurie plots of $\beta$ decays of nuclei as well as coherent low energy scattering of nuclei.
The latter has the added advantage that they can probe the new $\nu-\chi-\phi^0$ vertex,
while $\beta$ decay experiments do not.

In addition to producing a new gauge boson $Z_s$, the GP also allows $\chi$ to interact with the SM.
This allows $\chi$ to be a thermally produced dark matter candidate. The phenomenology of the $Z_s$, whose mass
is expected to be in TeV, can be probed in electroweak precision measurements \cite{CNW1} and directly searched
for at the LHC. The new invisible decay of the SM Z will also be an exciting channel for the Z-factory
option of future lepton colliders such as the FCC-ee and CEPC\cite{ CEPC, FCC}.

In this study, we also found a new role for the portal scalar $\phi^0$. If it has a mass in the GeV range,
it can act as an agent for entropy dilution and bring the relic density of the thermally produced
$\chi$ to the range of the observed value. It can also be looked for in precision measurements of the
Z-boson at the Z-factory\cite{CNW}.

Although the model we explored is self-contained and renormalizable, it suffers from the same hierarchy problem that plagues the SM because of the use of an elementary scalar. The $U(1)_s$ symmetry also has a Landau pole problem as in QED. Hence, it is reasonable to assume that it should be embedded in a more elaborate dark sector. If that is the case, one need not demand that $\chi$ is
a dark matter candidate. This opens up regions of parameter space that we have not discussed.
We reiterate that the region of parameter space we have studied was motivated by looking
at $\chi$ as a WDM candidate. The search of signatures for $\chi, Z_s$ and $\phi$ in
experiments that we have discussed should be conducted with this general setting in mind.

\begin{acknowledgments}
 WFC is supported by the Taiwan Ministry of Science and Technology under
Grant No. 106-2112-M-007-009-MY3.
TRIUMF receives federal funding via a contribution agreement with the National Research Council of Canada and the Natural Science and Engineering Research
Council of Canada.
\end{acknowledgments}

\subsection*{Note Added}
 After the paper was completed, we were informed that similar considerations
of new fermions production in coherent neutrino-nucleus scattering was
done  in \cite{Brdar}.  We agree with their results where they overlap with ours.
Ref \cite{Ballett:2019} considered a similar setup  with radiative neutrino mass generations.

\end{document}